\documentclass[letterpaper,12pt,notoc]{JHEP3}
\usepackage{graphicx}
\usepackage{amssymb}
\usepackage{amsmath}

\newcommand{\eeq}{\end{equation}}
\newcommand{\beq}{\begin{equation}}
\newcommand{\ba}{\begin{array}}
\newcommand{\ea}{\end{array}}
\newcommand{\bea}{\begin{eqnarray}}
\newcommand{\eea}{\end{eqnarray}}
\newcommand{\vev}[1]{\langle #1\rangle}
\newcommand{\vp}{\varphi}
\newcommand{\al}{\alpha}
\newcommand{\si}{\sigma}
\newcommand{\tha}{\theta}
\newcommand{\mat}[1]{{\mathbf #1}}

\title{
On the modular invariance of mass eigenstates and CP violation}
\author{Thomas Dent\thanks{tdent@umich.edu} \\
	{\em Michigan Center for Theoretical Physics, Randall Lab.,} \\
	{\em University of Michigan, Ann Arbor, MI 48109-1120}}
\date{November 2001}
\abstract{
We investigate the modular transformation properties of observable (light) 
fields in heterotic orbifolds, in the light of recent calculations of 
CP-violating quantities. Measurable quantities must be modular invariant 
functions of string moduli, even if the light fields are not invariant. We 
show that physical invariance may arise by patching smooth functions that 
are separately noninvariant. 
CP violation for $\vev{T}$ on the unit circle, which requires light and heavy 
states to mix under transformation, is allowed in principle, although the 
Jarlskog parameter $J_{CP}(T)$ must be amended relative to previous results. 
However, a toy model of modular invariant mass terms indicates that the 
assumption underlying these results is unrealistic. In general the mass 
eigenstate basis is manifestly modular invariant and coupling constants are 
smooth invariant functions of $T$, thus CP is unbroken on the unit circle. 
We also discuss the status of CP-odd quantities when CP is a discrete gauge 
symmetry, and point out a link with baryogenesis.}
\keywords{Compactification and String Models, Discrete and Finite Symmetries, CP violation}
\preprint{MCTP-01-50}

\begin{document}

\section{Introduction}
Recently, contradictory claims have appeared \cite{Lebedev,Dent01} concerning 
the invariance of 
physical quantities under T-duality transformation in heterotic orbifolds, 
particularly for CP-violating quantities such as the Jarlskog parameter, and 
the consequences for phenomenology. In the (somewhat simplified) models under 
discussion CP is assumed to be spontaneously broken by a $T$ (K{\" a}hler) 
modulus of compactification, which has the CP transformation $T\mapsto 
T^*$.\footnote{The imaginary part of $T$ is dual to the antisymmetric tensor 
field in the compact directions; thus its sign flips under orientation 
reversal of the compact space (equivalently, conjugation of the complex 
planes of the orbifold), which can be shown to be an appropriate CP 
transformation \cite{StromingerW,ChoiKN,DineLM,KobayashiL}.}  
Low-energy quantities such as Yukawa couplings and soft terms are functions 
of $T$ and inherit complex phases from its v.e.v.. The perturbative theory 
has a duality symmetry (properly, SL$(2,{\mathbb Z})$ modular symmetry) 
\cite{DijkgraafVV} to all orders, generated by 
\begin{equation}
{\mathcal T}\colon T\mapsto T+i,\ {\mathcal S}\colon T\mapsto 1/T,
\end{equation}
where the second generator ${\mathcal S}$, which exchanges large and small 
radii, is loosely 
referred to as T-duality. The action of the symmetry on orbifold twisted 
sectors \cite{LauerMN,LercheLW+FerraraLT} is unitary and intrinsically 
nonabelian: the generators cannot be simultaneously made diagonal by any 
constant change of basis. In the ``twist'' basis usually found convenient, 
the shift ${\mathcal T}\colon T\mapsto T+i$ is diagonalised.

Then we showed in \cite{Dent01} that given a unitary modular transformation 
of observable matter fields, CP was unbroken for modulus v.e.v.'s satisfying 
either $1/T = T^*$ or $T+i = T^*$, that is those for which the CP 
transformation is a modular transformation. In particular, if either Im$\,T=
\pm 1/2$ or $|T|=1$, this condition is satisfied 
and no CP violation is expected. In contrast, it was recently claimed in a 
calculation based on a $Z_6$ orbifold model \cite{Lebedev} that the Jarlskog 
parameter $J_{CP}(T)$ does not vanish on the unit circle, and explicit 
numerical results were presented. Here, the result for $J_{CP}(T)$ 
changes sign on taking $T\mapsto T^*$, as na\"\i vely expected for a CP-odd 
quantity. This claim raises an important question, whether observable 
quantities can depend on which one of two or more vacua related by 
SL$(2,{\mathbb Z})$ we live in. Since the theory is modular invariant, 
physics should be the same in vacua that are related by the modular group 
--- here, by $T\mapsto 1/T$. In fact, since CP can be shown to be part of the
gauge symmetry of the heterotic string \cite{ChoiKN,DineLM}, one also expects 
CP-conjugate vacua to give the same value (rather than values differing by a 
sign) for any measurable quantity, by the same argument: vacua differing only
by a constant element of the symmetry group are not physically distinguishable.

The calculation of \cite{Lebedev} was based on an important assumption 
which allows the previous result \cite{Dent01} to be evaded. In this paper 
we examine this and other assumptions about modular invariance and 
low-energy physics, and reach a possibly unexpected conclusion, based on a 
careful definition of physical observables: modular invariance may be restored
by patching smooth, noninvariant functions together, in a manner determined 
by which states are light at any particular point in moduli space. 
In principle, the result of \cite{Lebedev} may be correct for some 
modulus values, but must be amended for others to restore its modular 
invariance and physical significance.

In the second part of the paper we use a formally modular invariant toy 
model to test the conjecture that light and heavy matter can mix under 
modular transformation. Applying the assumption of \cite{Lebedev} to the 
model one obtains a somewhat unnatural result, requiring some mass matrix 
elements to vanish exactly at the physical value of $\vev{T}$. This condition 
is not maintained after modular transformation, and also results in 
an undesirable degenerate matter spectrum when $\vev{T}$ lies on the unit 
circle. In the more general case where all elements are nonzero, the rotation
to the mass eigenstate basis and the nonabelian modular transformations 
combine in a complicated fashion. 

Without plumbing the full algebraic depths
of the situation, we show by an appropriate choice of initial basis that our 
earlier guess \cite{Dent01}, that the mass eigenstates are invariant (up to 
unobservable phases) under the combined transformation of the modulus and the 
twisted states, was correct. Then since the action is invariant, all
$T$-dependent coupling constants (or rephasing-invariant combinations 
thereof) written in the mass basis are also modular invariant. Any 
non-invariant couplings in the original basis are ``killed'' by the 
{\em modulus-dependent}\/ change to the mass basis. This result should hold
also in more complex models, thus any observable property of the light matter
fields should be a unique, smooth, invariant function of $T$, and CP is 
unbroken for $T$ on the unit circle. 

We also reconcile the status of CP as a gauge symmetry with the existence 
of ``CP-odd'' observables, using some elementary thought experiments.

\section{Nonunitary quark transformations}
The technical point in \cite{Lebedev} that allowed the Jarlskog parameter 
$J_{CP}(T)$ as usually defined (as a function of quark mass matrices) not to 
be invariant under $T\mapsto 1/T$, is the transformation of the three 
generations of light quark fields.\footnote{Light meaning compared to the GUT 
or Planck scales.} The orbifold twisted sector in which the quarks live has 
many more states than required in the Standard Model. In general the states 
are labelled by the twists under which they are invariant and by the fixed 
points at which they are localized (see {\em e.g.}\/\ \cite{BailinL}). Then
in a given twisted sector the states mix unitarily under duality, as already 
stated. But if, as assumed in \cite{Lebedev} one picks a subset of these to 
represent the light quarks, the transformation of this subset will generally 
not be unitary, and will involve other states (assumed to be much heavier) in 
the twisted sector. After duality, the transformed Yukawa couplings or mass 
matrices are not related unitarily to the original couplings of only the 
light fields. 

We have, schematically,
\begin{equation}
{\mathcal S}\colon q_A \mapsto \tilde{q}_A\equiv 
\mat{S}_{AB} q_B,\ A,B=1,\dotsc,N_g \label{eq:qtransf}
\end{equation}
for the light and heavy generations together. The left-handed doublets and 
right-handed up- and down-type fields may each transform with a different 
$\mat{S}$, but to simplify the discussion (and also since transformations 
of the right-handed quarks cancel in the expression for $J_{CP}$) we consider 
a single matrix. Then for the light quarks $q_i$
\begin{equation}
q_i \mapsto \mat{S}_{ij} q_j + \mat{S}_{ia} q_a,\ i=1,2,3,\
a=1,\dotsc,N_g-3
\end{equation}
and the matrix $\mat{S}_{ij}$ will in general not be unitary (given that some 
$\mat{S}_{ia}$ are non-vanishing). Since the mass terms are invariant we 
have
\begin{equation}
{\mathcal S}\colon M^{u,d}_{AB}(T) \mapsto \tilde{M}^{u,d}_{AB}(T) \equiv 
M^{u,d}_{AB}(\tilde{T}) = (\mat{S}M^{u,d}(T)\mat{S}^\dag)_{AB}
\label{eq:MABtransf}
\end{equation}
and for the entries with the light state $q_i$ labels
\begin{multline}
M^{u,d}_{ij} \mapsto \tilde{M}^{u,d}_{ij} =
\mat{S}_{iA}M^{u,d}_{AB}(T)\mat{S}^\dag_{Bi} = 
\mat{S}_{ik}M^{u,d}_{km}(T)\mat{S}^\dag_{mj} \\
+ (\text{terms proportional to }M^{u,d}_{ab}(T)) \label{eq:Mijtransf}
\end{multline}
which again is generically not a unitary transformation and involves the 
mass matrix of the heavy states $q_a$.

Now the $J_{CP}$ parameter defined by 
\begin{multline} 
\det[M^u_{ij}M^{u\dag}_{jk},M^d_{mn}M^{d\dag}_{np}] = 
J_{CP}(m_t^2-m_c^2)(m_c^2-m_u^2)(m_t^2-m_u^2)\\
(m_b^2-m_s^2)(m_s^2-m_d^2)(m_b^2-m_d^2) \label{eq:Jdef1}
\end{multline}
or by
\begin{equation}
J_{CP} = {\rm Im}\,V_{11}V^*_{12}V_{22}V^*_{21}
\end{equation}
is invariant precisely under unitary changes of basis for the three quark 
generations \cite{Jarlskog+BernabeuBG}, which appear in the expression 
(\ref{eq:Jdef1}) as unitary 
transformations of the quark mass matrices. Thus with a nonunitary modular
transformation of quarks $J_{CP}(T)$, as defined on the mass matrix of the 
original states $M^{u,d}_{ij}$, may change in value. The corollary of this is 
that this value $J_{CP}(\tilde{T})$ represents some function of the couplings 
of, in general, a linear combination of light and unobservably heavy fields. 

In order to calculate the physically observed $J_{CP}$ in the modular 
transformed vacuum, the complete mass matrix of twisted fields must be 
block-diagonalised to find the three new light eigenstates, and the relevant 
Yukawa couplings extracted, resulting in a possibly different function 
$J'_{CP}(T)$; nevertheless, $J'_{CP}(\tilde{T})$ must have the same numerical 
value as $J_{CP}(T)$, since it is a physical observable dependent only on the 
modulus v.e.v., which cannot change under modular transformation of $T$ if
modular invariance is really a symmetry of the theory.\footnote{Similarly, in a GUT 
where some matter decouples ({\em e.g.}\/\ E$_6$), a physical measure of CP 
violation (or of anything else!) cannot change on exchanging one set of Higgs 
v.e.v.'s for a gauge-equivalent set; the full theory is not invariant under 
the transformation of the Higgses without also transforming the matter 
fields, but given the gauge-transformed Higgs v.e.v.'s one is forced
to use gauge-transformed matter fields, because they turn out to be the 
light fields.}

It was already assumed that the $q_i$ were the light states for the original
v.e.v., which we denote as $\vev{T}=T_0$: thus the mass matrices 
$M^{u,d}_{AB}(T_0)$ are block-diagonal in the $i,a$ basis. The light states 
in the transformed vacuum $\vev{T}=\tilde{T}_0$ result from diagonalising
the mass term 
\begin{equation}
\bar{q}_A M^{u,d}_{AB}(\tilde{T}_0) q_B = 
\bar{q}_A (\mat{S}M^{u,d}(T_0)\mat{S}^\dag)_{AB} q_B
\end{equation}
thus they are just $\psi_i=\mat{S}^\dag_{iA}q_A$. Then, trivially, the mass 
matrices from which $J'_{CP}$ is to be calculated are identical in value to 
the light quark mass matrices $M^{u,d}_{ij}$ in the original vacuum and we 
find (for any $T$)
\[ J'_{CP}(\tilde{T}) = J_{CP}(T). \]
Note that $J_{CP}(\tilde{T}_0)$ may in general be different from 
$J_{CP}(T_0)$: we have 
\begin{equation}
J_{CP}(\tilde{T}_0) \propto \det[\tilde{M}^u_{ij}\tilde{M}^{u\dag}_{jk},
\tilde{M}^d_{mn}\tilde{M}^{d\dag}_{np}]
\end{equation}
and the off-diagonal (light-heavy) transformations $\mat{S}_{iA}$ of 
(\ref{eq:Mijtransf}) do not cancel in the determinant. If one were to use the 
old function $J_{CP}$ at $\vev{T}=\tilde{T}_0$, as a function of the 
transformed mass matrices in the $q_i$ basis, such a result would apply to the 
linear superpositions $q_i=\mat{S}_{iA} \psi_A$, where the $\psi_A$ are both 
light and heavy fields, so the calculation is not physically meaningful.

Let us apply this to the model in which $J_{CP}(T)$ is claimed not to be 
invariant under the duality $T\mapsto 1/T$. On the unit circle the duality 
becomes $T\mapsto T^*$ and the lack of invariance allows a nonzero value of 
$J_{CP}$ (see Fig.~2 of \cite{Lebedev}). In the calculation, certain twisted
sector states are identified with the light quarks and others are assumed to 
be heavy. With this {\em ansatz}\/ the Yukawas and mass matrices are found, 
resulting in a smooth, non-modular-invariant function $J_{CP}(T)$ over the 
complex $T$ plane. Then the result may be correct over some of the unit 
circle (say, the part with Im$\,T>0$) but, since a nonunitary transformation 
of the kind discussed must be happening precisely between this line segment 
and its image under duality, the derived function $J_{CP}(T)$ cannot 
represent the physical quantity it is claimed to over the rest of the domain. 
It might seem {\em a priori}\/ unlikely that a specific mechanism would make 
precisely those states light which are picked out in this calculation: given 
a specific mass matrix for the whole twisted sector, the light states would 
most probably be mixtures of the states in the fixed point basis. But granted
the assumptions of \cite{Lebedev}, some light states must mix with heavy 
ones under modular transformation in order for $J_{CP}(T)$ to be nonzero
on the unit circle, so the derived function $J_{CP}(T)$ does not remain 
physical after modular transformation. 

Instead we need a new function $J'_{CP}(T)$ valid over the rest of the domain
as described above. Since the duality is ${\mathbb Z}_2$, we have
$J'_{CP}(T)\equiv J_{CP}(\tilde{T})$ and $J'_{CP}(\tilde{T})\equiv J_{CP}(T)$, 
and realising that the original function $J_{CP}(T)$ is odd under complex 
conjugation we find that $J'_{CP}(T)= -J_{CP}(T)$ must hold on the unit 
circle. Thus the physical Jarlskog parameter is given by $\pm J_{CP}(T)$, the 
sign depending on which domain $\vev{T}$ lies in. In particular, there will be 
a ``join'' along which the two prescriptions collide, and the physical result 
will not in general be differentiable at that point. We know we are using the 
``right'' prescription precisely when the couplings that appear in the 
calculation are just those of the light fields. In the $Z_6$ orbifold model, 
one should find the explicit mechanism that makes the heavy fields decouple: 
the expectation is that different (linear combinations of) fields will be 
the light ones over different regions of moduli space. The manifestly modular 
invariant result on the unit circle will take the form 
$J^{\rm (phys)}_{CP}=\pm|J_{CP}(T)|$, the overall sign depending on which 
way the prescription turns out (see Fig.~\ref{fig:corr} below).

There is a further sign ambiguity in this (and any theoretical) result for a 
CP-violating quantity, connected with the definition of matter and antimatter;
we return to this at the end of the paper.

The result of \cite{Lebedev} is already invariant under the axionic shift 
generator $T\mapsto T+i$, and one might ask why this should be, since in a 
generic basis of states (such as the basis of light and heavy eigenstates 
will likely be) both generators will be off-diagonal. The reason is simply 
that the basis of light and heavy states was assumed to be exactly that in 
which $T\mapsto T+i$ did act as a diagonal matrix, thus the argument of 
\cite{Dent01} applies to this symmetry and CP symmetry is unbroken at 
Im$\,T=\pm 1/2$. 

We briefly return to the result of \cite{Dent01} and ask, what went wrong
with our reasoning for the $T\mapsto 1/T$ transformation? Unitary quark 
transformations were not strictly necessary for our result, but we required 
that a CP transformation acting on observable fields, followed by the action 
of a modular transformation on those fields, remained a physically reasonable 
CP transformation. But if there is mixing of light and heavy matter fields 
under duality, the resulting ``general CP transformation'' cannot correspond 
to anything testable in the laboratory: light particles have to remain light 
under CP.\footnote{Within the latitude implied by the CKM mixing, which allows 
us to mix, for example, bottom with down.} However, if we imagine performing 
experiments with the heavy twisted states and their antiparticles, some kind of 
unbroken CP symmetry may reappear for $\vev{T}$ lying on the unit circle.

\section{A simple illustration}
Since the nonabelian and nonunitary transformations involved in the system 
under consideration are rather complicated, we illustrate the general 
principle with a simple gedankenexperiment. Suppose we have deduced the 
correct value of the string scale $M_s$ indirectly, and suspect that we live 
in a ``large-radius'' compactification (that is, with some radii differing 
significantly from $M_s^{-1}$). Given $T$-duality, we could equally well call 
it a ``small-radius'' compactification. Then imagine an accelerator 
experiment that could probe the energy of the lowest K-K or winding mode (but 
not the string scale itself). Clearly we cannot say that the mass of the mode 
that we hope to measure is given unambiguously in string units by $1/R$ (for 
a K-K mode) or $R$ (winding mode). But there is a rather trivial and 
manifestly duality invariant expression for this mass, namely
\begin{equation}
m_{\rm detected} = \min (R, 1/R)
\end{equation}
in other words two noninvariant, continuous functions over moduli space, 
with a prescription for choosing between them. The prescription has a clear 
physical interpretation, namely that the lighter mode will be found at the 
experiment. The result is nonanalytic and not differentiable at $R=1$, 
just as the prescription for $J^{\rm (phys)}_{CP}(T)$ 
is at the real axis. Conversely, extending the smooth function $J_{CP}(T)$ 
calculated in \cite{Lebedev} over a region containing both $T_0$ and 
$\tilde{T}_0$ would correspond to assuming that one smooth function, say 
$m_{\rm detected}=R$, holds for all values of $R$. 

Of course, considering also the string excited states with mass scale $1$ in 
our units, the real behaviour is likely to be more complicated near $T=1$, 
but in the approximation that only the ``geometric'' (winding/KK) modes 
contribute our prescription appears reasonable.

\section{The fundamental domain}
One strategy to deal with the problem of degenerate vacua in a modular 
invariant theory is to restrict the value of $T$ to the fundamental domain 
$\mathcal{F}$, usually defined as the region satisfying 
$-1/2<{\rm Im}\,T\leq 1/2$, $|T|\geq 1$. 
The procedure is analogous to gauge-fixing. Any value in the half-plane 
Re$\,T>0$ can be reached by a modular transformation from exactly one point 
in $\mathcal{F}$ --- no new physics occurs when we go outside. If we are 
interested in values on the unit circle then one half of the line segment 
$\mathcal{P}$ between $e^{-i\pi/6}$ and $e^{i\pi/6}$ must be excluded: let 
us keep the half with Im$\,T\geq 0$. Then if we want to find the value of 
$J_{CP}$ for $T$ lying on the excluded part of $\mathcal{P}$, we just 
calculate it at the modular-equivalent value $T^*=1/T$ lying in $\mathcal{F}$. 
On this basis, at first glance there is no objection to the result of Fig.~2 
of \cite{Lebedev}, as long as half of the graph is ignored as being outside 
$\mathcal{F}$, and replaced by the mirror reflection of the other half in 
$\phi=0$ (Fig.~\ref{fig:corr}). 
But this replacement creates trouble for the 
rest of $\mathcal{F}$ away from $|T|=1$: since the original, noninvariant 
function $J_{CP}(T)$ was continuous over the half-plane, the change of sign 
would create a discontinuity akin to a branch cut along the excluded part of 
$\mathcal{P}$. Clearly the replacement along $\mathcal{P}$ is not enough, 
since a discontinuity is definitely unphysical. 
\FIGURE{
\includegraphics[width=11cm,height=8.5cm]{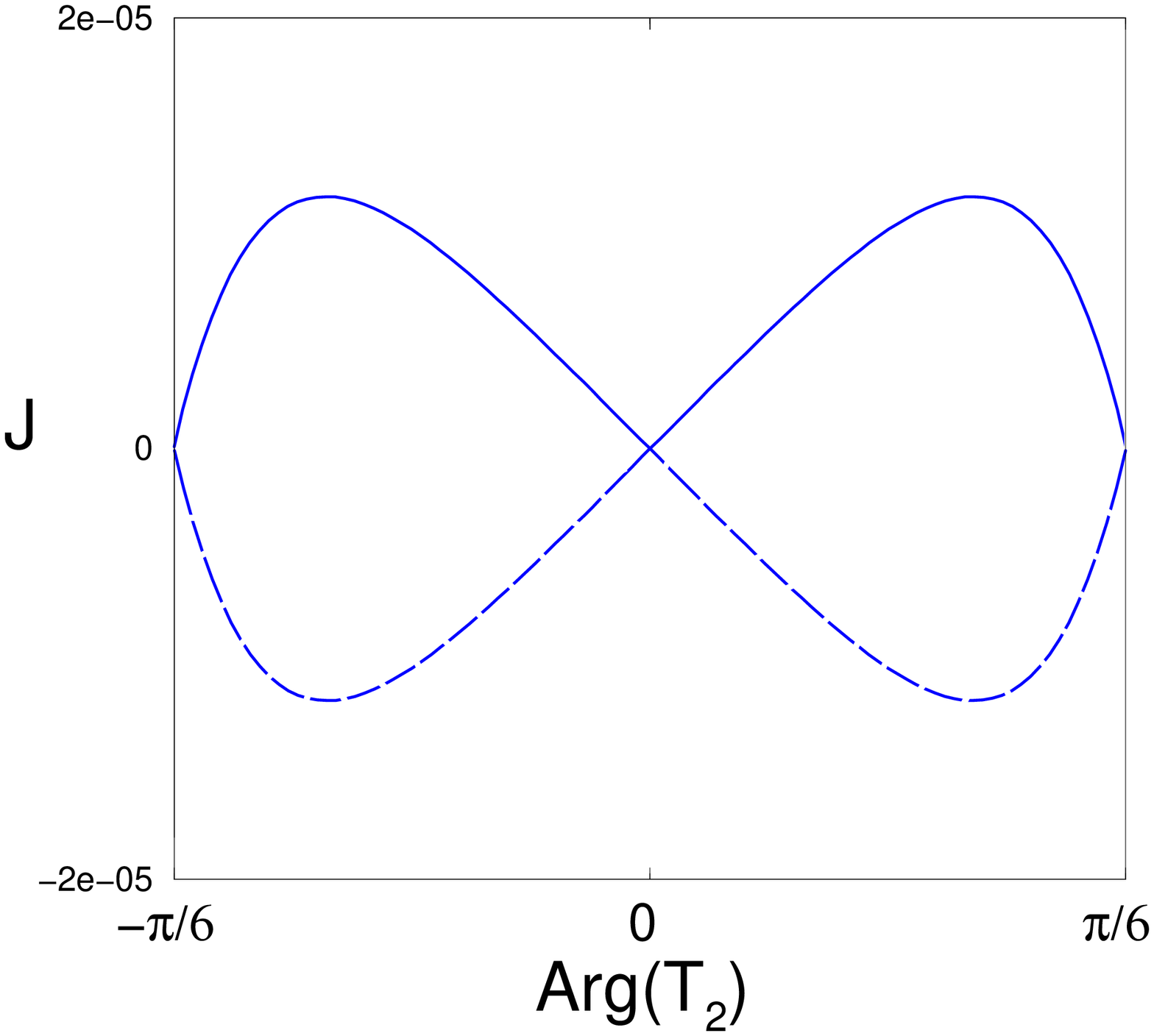}
\caption{Amended version of Fig.~2 of \cite{Lebedev}: the physical $J_{CP}$
parameter, invariant under duality and CP transformations of the modulus, is 
given by either the positive (solid) or the negative (dashed) value, 
depending on the correct identification of the light quark fields and 
of matter {\em vs.}\/\ antimatter.}
\label{fig:corr}}

Thus the smooth function $J_{CP}(T)$ must be amended over a whole region of 
$\mathcal{F}$. In fact, taking modular invariance in conjunction with the 
exact CP symmetry of the underlying theory (discussed in section 
\ref{sec:CPsymm}), we find that the physical value of $J_{CP}$ must be 
invariant under complex conjugation of $T$. Thus the most likely possibility 
is that the ``kink'' where the two prescriptions join is on the real axis, in 
which case half of the fundamental domain of SL$(2,{\mathbb Z})$ needs 
replacement, and our previous answer $J^{\rm (phys)}_{CP}=\pm |J_{CP}(T)|$ is 
extended to the whole of $\mathcal{F}$. Given that CP is also
an exact symmetry acting on $T$, it makes sense to define a fundamental 
domain $\mathcal{F}'$ for the group SL$(2,{\mathbb Z})\otimes CP$, which can 
be taken to be the half of $\mathcal{F}$ with Im$\,T\geq 0$: then any
value of $T$ is physically equivalent to precisely one point in $\mathcal{F}'$. 
If we only consider $T$ inside $\mathcal{F}'$ then there is no need to patch 
noninvariant functions.

What happens near $T=1$? If, as needed for nonunitary transformation of the 
light quarks, light and heavy states mix under duality, and mass matrix 
entries are continuous functions on the moduli space, something odd happens on 
passing through the self-dual point: light and heavy states cross over. We 
show in the next section that if the original basis was the mass basis, 
implying that off-diagonal entries vanish in the mass matrix, the self-duality
condition which holds at $T=1$ leads to an unacceptable degenerate mass 
spectrum. Thus the theory near this point bears no relation to the Standard 
Model and no value of $J_{CP}$ can be meaningful. This may not count 
significantly against the assumptions of the model, since there is anyway no 
CP violation at $T=1$. However, a similar argument also applies for values 
of T approaching the unit circle: the combination of $T\mapsto 1/T$ and CP 
transformation brings us to a physically equivalent v.e.v.\ just on the other 
side of the circle, but also (by assumption) mixes the light and heavy states. 
Thus a degenerate spectrum also occurs on the ``CP self-dual'' line $|T|=1$. If,
however, the mass matrix is not diagonal in the original basis, the spectrum 
need not be degenerate at these values of $T$; but then we find that the modular 
transformation of the resulting mass eigenstates is trivial.

It is of course also possible in this class of theories that the light quarks 
{\em do}\/ transform unitarily among themselves, in other words that there
are no (light-heavy) off-diagonal elements in the representation of the 
modular group. In this case $J_{CP}(T)$ is a unique modular invariant smooth 
function odd under $T\mapsto T^*$, which must vanish on the boundary of 
$\mathcal{F}$, and no prescriptions are needed to maintain modular invariance. 
But considerations of CP symmetry, discussed above and in more detail in 
section \ref{sec:CPsymm}, may even in this case force us to adjust the result.

\section{Modular invariance and mass eigenstates: a toy model}
\label{sec:masses}
In order to see in more detail what happens at a self-dual point and what 
behaviour of light and heavy matter fields is consistent with a modular 
invariant theory, we consider a 2-by-2 mass matrix for complex scalars. This 
is to be our toy model for the $N_g-3$ heavy and $3$ light states that are 
supposed to result from a more realistic orbifold-based construction. We 
consider a nonabelian duality group with two generators, and formally write the
modular invariant mass term as 
\begin{equation} 
\mat{M} = \left(\ba{ccc} a &b \\ b^* &d \ea\right)
\end{equation}
where $a$, $b$, $d$ are functions of the modulus, with transformations under
the duality group to be determined. We take a group action similar to the 
representation of SL$(2,{\mathbb Z})$ on the twist fields $\sigma_i$, $i=1,2,3$, 
of the two-dimensional orbifold $T^2/Z_3$, for which
\begin{equation}
{\mathcal S}=-\frac{i}{\sqrt{3}}\begin{pmatrix} 1 &1 &1 \\ 1 &\al &\al^2 \\ 
1 &\al^2 &\al \end{pmatrix},\
{\mathcal T}=\begin{pmatrix} \al &0 &0 \\ 0 &1 &0 \\ 0 &0 &1 \end{pmatrix}
\end{equation}
where $\al=e^{2\pi i/3}$ \cite{LercheLW+FerraraLT}.\footnote{There is a sign
error in the ${\mathcal S}$ generator in the first reference of 
\cite{LercheLW+FerraraLT}.} This representation is reducible into a singlet 
$\tau_0=\sqrt{2}^{-1}
(\sigma_2-\sigma_3)$ and a doublet
\bea 
\tau_1 &\equiv& \si_1 \nonumber \\ 
\tau_2 &\equiv& \frac{1}{\sqrt{2}}(\si_2+\si_3) 
\eea
for which 
\begin{equation}
{\mathcal S}^{(\tau)}=-\frac{i}{\sqrt{3}}
\left(\ba{cc} 1 &\sqrt{2} \\ \sqrt{2}&-1 \ea\right),\ 
{\mathcal T}^{(\tau)}=
\left(\ba{cc} \al &0 \\ 0 &1 \end{array}\right).
\end{equation}
We can generalise the representation slightly without complicating the system
by taking
\begin{equation}
{\mathcal S}^{(\tau)}= -i\begin{pmatrix} 
\cos\tha_S &\sin\tha_S \\ \sin\tha_S &-\cos\tha_S \end{pmatrix},\ 
{\mathcal T}^{(\tau)}= \begin{pmatrix}
\al &0 \\ 0 &1 \end{pmatrix}. \label{eq:tautransf}
\end{equation}
The modular invariance conditions for the mass matrix analogous to 
(\ref{eq:MABtransf}) are
\begin{equation*}
{\mathcal S}\colon \mat{M} \mapsto \tilde{\mat{M}} = 
\left(\ba{ccc} \tilde{a} &\tilde{b} \\ \tilde{b}^* &\tilde{d} \ea\right),\
{\mathcal T}\colon \mat{M} \mapsto \bar{\mat{M}} = 
\left(\ba{ccc} a &\bar{b} \\ \bar{b}^* &d \ea\right)
\end{equation*} 
where
\begin{equation}
\begin{split}
\tilde{a}(T)& = a c_S^2 + d s_S^2 + 2{\rm Re}\,b c_S s_S, \\
\tilde{b}(T)& = (a-d)c_S s_S - {\rm Re}\,b(c_S^2-s_S^2) - i{\rm Im}\,b, \\
\tilde{d}(T)& = a s_S^2 + d c_S^2 + 2{\rm Re}\,b c_S s_S, \\
\bar{b}(T)& = \rho b.
\end{split}
\end{equation}
with $c_S=\cos\tha_S,$ $s_S=\sin\tha_S$.

The nonunitary transformation required for CP violation on the unit circle 
corresponds to nonzero off-diagonal ({\em i.e.}\/\ not block diagonal) elements 
in the (light, heavy) basis, for one or more group generators. In the toy model 
this will be diagnosed by $(1,2)$ and $(2,1)$ entries for the 
transformations in the mass basis, or by any change 
under modular transformation in the linear superposition of the original 
$\tau$ making up the mass eigenstates. The (modular invariant) mass 
eigenvalues are $m^2_{1,2}=\frac{1}{2}(a+d\pm\sqrt{(a-d)^2+4|b|^2})$.

Without further calculation we can see the effect of imposing a self-duality 
condition corresponding to the behaviour of quark mass matrices at the point 
$T=1$. At $T=1$ we have ${\mathcal S}(T)=T$, thus $\tilde{\mat{M}}=\mat{M}$. 
From this it can be deduced that
\begin{equation}
a-d = 2{\rm Re}\,b \cot\tha_S,\ {\rm Im}\,b = 0 \label{eq:selfdual}
\end{equation}
and the mass eigenvalues become $\frac{1}{2}(a+d \pm\sqrt{(a-d)\sec\tha_S})$. 
Because of the mixing in going from the 
$\tau$ basis to the mass basis the mass spectrum can be nondegenerate, 
although to achieve the desired hierarchy $m_2\ll m_1$ fine-tuning seems to be
required. If we literally implement the assumption of \cite{Lebedev}
that the $\tau_i$ are mass eigenstates, thus $b=0$, we find an unacceptable
spectrum at $T=1$: the self-duality relation (\ref{eq:selfdual}) reduces to 
$a=d$ and the states are degenerate. 

This value is anyway uninteresting for CP violation; to model the case of 
$T$ on the unit circle we need to implement the ``CP-self-duality'' relation 
$CP(T)=T^*={\mathcal S}(T)$ on the mass matrix entries. We require 
$\tilde{\mat{M}}=\mat{M}^*$ \footnote{One may also introduce diagonal matrices 
of constant phases in the CP transformation of $\mat{M}$, corresponding to the 
$\tau_i$ receiving complex phases under CP; this possibility does not affect
our conclusions.} which reduces to the single condition
\begin{equation}
a-d = 2{\rm Re}\,b \cot\tha_S
\end{equation}
with no restriction on Im$\,b$. Thus the mass eigenvalues in the case where 
the $\tau$ are not mass eigenstates are given by three parameters $a$, $d$ 
and Im$\,b$ and there is more freedom to obtain a large hierarchy. But in the 
case that $\tau_i$ are mass eigenstates (so $b=0$) the spectrum is still 
degenerate.

Thus, the possibilities that $\vev{T}$ lies inside ${\mathcal F}$ with the 
$\tau_i$ being mass eigenstates, or that $\vev{T}$ lies on the unit circle
with the mass basis distinct from the $\tau$ basis, remain viable, but we 
cannot live in a world where the $\tau_i$ are mass eigenstates and $|T|=1$ 
(regardless of the status of CP). Note also that if $b=0$ holds at one value 
of $T$ it cannot hold at the dual value $\tilde{T}$ since 
$\tilde{b}=(a-d)s_Sc_S$, so one cannot consistently impose $b=0$ over all of
moduli space unless $\tha_S=\pm \pi$. Unless the light and heavy states 
are {\em permuted} with no mixing under ${\mathcal S}$ (which is inconsistent 
with any known behaviour of twisted states), or we happen to live at the rather 
special point(s) in moduli space where the non-modular-invariant condition 
$b=0$ holds, the mass eigenstates (modelling the three light and $N_g-3$ heavy 
generations) are unitary mixtures of the twist states. 

Then the modular properties of the mass eigenstates have to be computed 
explicitly. We consider transformation of both the modulus and the $\tau$
fields together, since this is the symmetry under which the theory is
invariant, thus the properties of the theory written in the mass eigenstate 
basis will be easy to find. We have
\begin{equation}
\tau^\dag \mat{M} \tau = \phi^\dag {\rm diag}(m^2_i) \phi = 
\phi^\dag \mat{U}^\dag \mat{M}\mat{U} \phi
\end{equation}
where $\mat{U}$ is given by
\begin{equation}
\mat{U} = e^{i\zeta}\left(\ba{cc} \cos\tha &\sin\tha e^{i\vp} \\ 
-\sin\tha e^{i\chi} & \cos\tha e^{i(\vp+\chi)} \ea \right) \label{eq:Udef}
\end{equation}
with parameters 
\begin{equation}
\chi = -\arg b,\ \tan 2\tha = -\frac{2|b|}{a-d}.
\end{equation}
The phases $e^{i\zeta}$ and $e^{i\vp}$ are arbitrary. The transformation of 
the $\phi$ under ${\mathcal T}$ is simply
\begin{equation}
\phi_i \mapsto \bar{\phi} = \rho \phi_i 
\end{equation}
since the transformation of $b$ cancels against that of $e^{i\chi}$. This 
result can be further reduced to $\bar{\phi}=\phi$ by an appropriate choice 
of $\zeta$. On the other hand the ${\mathcal S}$-transformed eigenstates 
$\tilde{\phi}$ are more complicated functions of the $\tilde{\tau}$, which 
themselves are mixed relative to the $\tau$ under ${\mathcal S}$:
\begin{multline}
\tilde{\phi}=\tilde{\mat{U}}^\dag\tilde{\tau}, \tilde{\chi} = 
-\arg(\tfrac{a-d}{2}\sin 2\tha_S - {\rm Re}\,b\cos 2\tha_S - i{\rm Im}\,b), \\
\tan 2\tilde{\tha} = -\frac{2|\tfrac{a-d}{2}\sin 2\tha_S - 
{\rm Re}\,b\cos 2\tha_S - i{\rm Im}\,b|}{(a-d)\cos 2\tha_S - 
2{\rm Re}\,b\sin 2\tha_S}. \label{eq:phitilde}
\end{multline}
The simplification of these formulae is scarcely practicable given that one
requires $\cos\tilde{\tha}$, $\sin\tilde{\tha}$ to be found explicitly in 
order to write $\tilde{\phi}$ in terms of $\tau$.

Algebraic difficulties are however easily circumvented by a constant change of 
basis of the original states. Since any unitary matrix is unitarily similar
to a diagonal matrix, one can always find a basis 
$\varsigma = \mat{V}^\dag\tau$ in which any given modular transformation, 
for example ${\mathcal S}$, is diagonal. The transformation in the new basis 
is derived as 
\begin{equation} 
\mat{V}^\dag \tau \mapsto \mat{V}^\dag \mat{S}^{(\tau)} \tau
\end{equation}
(applying the constant linear combination $\mat{V}^\dag$ to $\tau \mapsto 
\mat{S}^{(\tau)} \tau)$
\begin{equation} 
\Rightarrow \varsigma \mapsto \mat{V}^\dag \mat{S}^{(\tau)} \mat{V} \varsigma
\equiv \mat{S}^{(\varsigma)} \varsigma
\end{equation}
(substituting for $\tau$ on the RHS). The condition
$\mat{S}^{(\varsigma)}={\rm diag}(e^{i \beta_1},e^{i \beta_2})$ is
easily solved by 
\begin{equation} 
\mat{V} = \begin{pmatrix} \cos\tha_V &\sin\tha_V \\
-\sin\tha_V &\cos\tha_V \end{pmatrix}
\end{equation}
with $\tha_V=-\tha_S/2$ and $\beta_1=-\pi$, $\beta_2=\pi$.\footnote{For a more general form of $\mat{S}^{(\tau)}$ complex phases may be needed in $\mat{V}$.}
The mass eigenstates $\phi$ are written
\begin{equation}
\phi={\mat{U}'}^\dag\varsigma
\end{equation}
where $\mat{U}'$ diagonalises the mass matrix $\mat{M}'$ in the $\varsigma$ 
basis. The parameters of $\mat{U}'$, defined analogously to (\ref{eq:Udef}), 
can be found in terms of the original 
mass matrix entries $a,b,d$; the expressions are similar to those for 
$\tilde{\chi}$ and $\tilde{\tha}$ in (\ref{eq:phitilde}) and equally 
unenlightening.\footnote{They turn out to be the same as in 
(\ref{eq:phitilde}) except for a sign and the substitution of $2\tha_S$ by 
$\tha_S$.} But since the $\varsigma_i$ have the simple transformation law
${\mathcal S}\colon \varsigma_1\mapsto -i\varsigma_1,\
\varsigma_2\mapsto i\varsigma_2$ it is easy to check that the transformed
diagonalisation matrix $\tilde{\mat{M}}'$ is the same as $\mat{M}'$ up to
a change of sign of $e^{i \chi'}$. Then under transformation of $\mat{M}'$
and $\varsigma$ together the $\phi_i$ just get an unobservable common factor 
$-i$, or with an appropriate choice of $\zeta'$ can be made invariant.

This result is to be expected since the ${\mathcal S}$ transformation in the
$\varsigma$ basis is exactly analogous to ${\mathcal T}$ acting in the $\tau$
basis. It is not difficult to show for a representation of arbitrary 
dimension that one can always find a basis $\sigma$ in which any given 
transformation is a diagonal matrix of phases. Then a modulus-dependent 
matrix analogous to $\mat{U}'$ can always be found such that the mass 
eigenstates are exactly invariant under transformation of both the modulus and 
the $\sigma_i$ (see Appendix). Then if the theory is written in the mass basis, 
all modulus-dependent coupling constants must also be exactly modular invariant
smooth functions. 
This situation was forecast in \cite{Dent01} as a solution to 
the problem of non-invariant coupling constants. Applied to the case of 
$\vev{T}$ on the unit circle it means that any apparently CP-violating
couplings calculated in the original basis of the theory must be killed by the 
modulus-dependent transformation to the basis of light and heavy states. The 
only exceptions are if the original basis happens precisely to be the mass 
basis, such that no diagonalisation is needed, or if the whole modular group 
acts either diagonally ({\em i.e.}\/\ trivially) or by a pure permutation (as
in the case of KK/winding modes). We argued that the coincidence of the theory
basis with the mass basis is highly unlikely, since such a condition cannot be 
imposed over any extended region of moduli space; also, no known examples of
the modular group action on twisted states are purely diagonal or permutation
matrices \cite{LauerMN}.

\section{CP as a gauge symmetry and ``CP-odd'' quantities} \label{sec:CPsymm}
Just as for modular invariance, we expect that no physically measurable 
quantity should allow us to differentiate between CP-conjugate vacua, since 
CP is an exact symmetry in the underlying higher-dimensional theory, 
being embeddable in the gauge group \cite{ChoiKN,DineLM}. 
This sounds odd at first: surely in the CP-conjugate vacuum, the CKM phase 
would have the opposite sign and the vertex $(\rho ,\eta)$ of the unitarity 
triangle would live in the lower half of the complex plane --- with easily 
measurable consequences? But recall that the exact CP symmetry involves 
conjugating both the scalar v.e.v.\ and the particle excitations. Then, 
imagining a ``CP domain wall'', what gives the same physics on the other side 
of the wall from us is a world with matter and antimatter exchanged. 
Combining CP with SL$(2,{\mathbb Z})$ we obtain a fundamental domain half the 
size of ${\mathcal F}$, with a unique sign for Im$\,T$: the ``kink'' at the 
real axis is now at the {\em edge}\/ of the domain.

The gauge aspect of the symmetry was exploited in \cite{ChoiKN} where ``CP 
strings'' were proposed, in the phase where the scalar v.e.v.\ still left CP 
intact (see also \cite{Schwarz}). In the broken phase, a traveller could 
conceivably go round a closed loop in space and pass through exactly one ``CP 
domain wall'', since the two vacua are gauge-equivalent and can be identified 
(the domain wall ending on a string). But the traveller ``around the string'' 
would, on returning, appear to be made of antimatter; or, deeming himself 
still matter, he would measure the opposite sign for CP asymmetries, and 
think that the scalar v.e.v.\ had flipped (on passing through the wall). A 
more sophisticated traveller might even, on recognising his (her?) 
surroundings again (and deducing that the v.e.v.\ was the same as at the 
outset) but getting the ``wrong'' sign for CP asymmetry experiments on board 
ship, decide that predictions should now be made with Hermitian conjugate 
fields (henceforth called antifields) replacing fields, thus restoring 
predictive power to his (her) favourite theory.\footnote{The astute deduction 
that he (she) was now made of antimatter would no doubt forestall the usual 
catastrophic end to such journey.} 

In order to predict a CP-violating quantity in such a theory, we need to 
establish a convention for matter {\em vs.}\/\ antimatter. Otherwise we face a 
sign ambiguity, since one cannot {\em a priori}\/ decide whether the fields or 
the antifields of the theory are to describe the experimental apparatus. This 
corresponds to a further $\pm$ sign in front of our modular invariant prescription 
for $J_{CP}$, which (temporarily) restores the symmetry between the positive and 
negative values in Fig.~\ref{fig:corr}. In the light of the traveller's 
predicament, it may not be possible to establish a globally consistent convention, 
but locally in regions with a net density of (say) baryon number, it is unambiguous 
and historically inevitable to pick the prevailing species as matter.\footnote{The 
alternative is to take an established experimental result and compare with the 
theory to fix the convention, so that only relative signs between different 
measurements can be predicted.} 

Little attention has been paid to the possible influence of ``CP strings'' on 
baryogenesis, perhaps not surprisingly as they erase the distinction between 
matter and antimatter in their neighbourhood. The usual assumption, valid if 
symmetry is broken at a high enough scale, is that all strings and domain walls 
have ``cleared out'' of the observable Universe, in which case baryogenesis 
proceeds uniformly over the observable region. This will in the end definitely 
establish one sign convention. On the other side of a putative CP domain wall, 
both the baryon asymmetry and the CP-violating v.e.v.\ would be conjugated, and 
{\em experimental results would be the same}\/.

Since such walls are cosmologically excluded, these speculations would 
seem to be irrelevant. But the point should be addressed in order to predict CP 
violation at experiments: the status of matter and antimatter in one's theory 
cannot be set by {\em fiat}\/, but must be deduced from a theory of 
baryogenesis. Otherwise one does not know (even in the correct theory) whether 
to calculate $J_{CP}$ from the quark mass matrices or from their Hermitian 
conjugates. 

\section{Conclusions}
In this paper we investigated the consequences of an exact discrete 
non-Abelian symmetry acting on moduli and matter fields. We were motivated
by the claim that the symmetry could be realised in a way that allowed light
and heavy states to be mixed, which can lead to interesting consequences for
phenomenology. For example, a nonvanishing CKM phase for $\vev{T}$ on the unit
circle would be desirable since such values appear to arise naturally from
gaugino condensation models of moduli stabilization and result in an exactly
vanishing modulus $F$-term, which might be part of a solution of the 
supersymmetric CP and flavour problems.

We argued that the light-heavy mixing, which leads to noninvariant 
results for observable moduli-dependent quantities, was possible in principle,
but that invariance must be restored by patching together more than one such 
function, the choice being made by correctly identifying the light fields and 
using the symmetries of the theory. The result may be a nondifferentiable 
function on moduli space.

In the second part of the paper we looked more carefully at what mass matrices
could be consistent with the assumption of light-heavy mixing. We found that 
the {\em ansatz}\/ of \cite{Lebedev}, in which the original basis of twisted 
states is also the mass basis, cannot produce a realistic spectrum at points
in moduli space invariant under $T\mapsto 1/T$ or $T\mapsto 1/T^*$. Even at other
values of $T$, this {\em ansatz}\/ is not preserved under modular 
transformation, so cannot hold except at isolated points and appears unnatural. 

A more general {\em ansatz}\/ with no vanishing mass matrix elements is
consistent with a spectrum of light and heavy states after diagonalisation,
but we find that these states are separately modular {\em invariant}\/, due 
to the transformation of the diagonalisation matrix. This simple result is 
significant since it implies that (measurable combinations of) the coupling 
constants of the light fields are exactly modular invariant functions.

Lastly, we discussed the implications of taking the exact CP symmetry of the 
underlying theory seriously, set against results which apparently imply 
different predictions in CP-conjugate vacua. A result such as Fig.~2 of 
\cite{Lebedev} for a physical measure of CP violation tends to produce the 
impression that, since the two vacua give opposite signs for CP-violating 
observables and we can't predict which vacuum we live in, the best we can do 
is predict the magnitude. But, after we realise that the value of such an 
observable cannot depend on the choice between (CP and modular) equivalent 
vacua and arrive at the amended result of Fig.~\ref{fig:corr}, we find 
that the sign can be predicted, but only in conjunction with the prediction 
of a nonzero baryon fraction.\footnote{Unfortunately for 
phenomenologists, it is conceivable that the source of CP violation at 
baryogenesis is unobservably small at the present epoch, and independent of 
the source of currently observed effects. In this case, which may be 
realised in Affleck-Dine-type ``spontaneous baryogenesis'' models, only 
relative signs between different experimental results can be predicted.} Once 
this is done the conjugate vacua give the same physics, as they should. 
Unfortunately the Jarlskog invariant is just what cannot tell us anything 
about baryogenesis, since generating the observed baryon asymmetry requires
CP violation in physics beyond the Standard Model \cite{GavelaHOPQ+HuetS}.

\acknowledgments
The author would like to thank David Bailin for motivating this study of 
modular invariance, Steve Thomas for suggesting the toy model investigation, 
Oleg Lebedev for an enlightening correspondence and Marty Einhorn for 
discussions. 
Research supported in part by DOE Grant DE-FG02-95ER40899 Task G.

\appendix
\section{Derivation of the modular invariance of mass eigenstates}
The invariance of the mass eigenstate basis is not manifest for the complete
modular group simultaneously, hence we treat each group element $\Gamma$, acting
as a unitary matrix on properly-normalised matter states, separately. By a constant
$\Gamma$-dependent change of basis one may always find a basis $\nu$ where $\Gamma$ 
acts as a diagonal matrix of phases:
\begin{equation}
\Gamma\colon \nu_{(L,R)a} \mapsto e^{i\xi_{(L,R)a}} \nu_{(L,R)a}
\end{equation}
where for greater generality we allow $\nu$ to range over left- and right-handed 
fermions. Then the mass matrix $\mat{M}^{(\nu)}$ in this basis is diagonalised as
\begin{equation}
\bar{\nu}_L \mat{M}^{(\nu)} \nu_R = \bar{\psi}_L {\rm diag}(m_i) \psi_R
\end{equation}
with $m_i$ real, where
\begin{equation}
\nu_{(L,R)} = \mat{U}_{(L,R)}\psi_{(L,R)},\ \mat{U}_L^\dag\mat{M}^{(\nu)}\mat{U}_R=
{\rm diag}(m_i)
\end{equation}
where $\mat{U}_{(L,R)}$ are modulus-dependent unitary matrices. The mass matrix 
changes under the modular transformation $\Gamma$ as
\begin{equation}
M^{(\nu)}_{ab} \mapsto \tilde{M}^{(\nu)}_{ab} = e^{i(\xi_{La}-\xi_{Rb})} 
M^{(\nu)}_{ab}
\end{equation}
thus the diagonalisation matrices should transform as
\begin{equation}
U_{Lai} \mapsto \tilde{U}_{Lai} = e^{i\xi_{La}}U_{Lai},\ U_{Rai} \mapsto
\tilde{U}_{Rai} = e^{i\xi_{Ra}} U_{Rai}
\end{equation}
and the transformed mass eigenstates are
\begin{equation}
\tilde{\psi}_{(L,R)i} = \tilde{U}^\dag_{(L,R)ia}e^{i\xi_{(L,R)a}}\nu_{(L,R)a}
= \psi_{(L,R)_i}
\end{equation}
formally demonstrating the invariance. This result is significantly different from 
the case of spontaneously broken continuous symmetry: in that case one can usually 
redefine the scalar v.e.v.\ to obtain a mass matrix diagonal in the symmetry
space ({\em e.g.}\/\ imposing $\vev{H}=(0,|v|/\sqrt{2})$ in the Standard Model 
using SU$(2)$ symmetry) and the mass eigenstates can be thought of as charged under 
the nonabelian group.
When considering string models with heavy matter ($\gg m_t$) one may 
``block-diagonalise'' the quark mass matrix by the above procedure, to separate the 
three light states from the rest, but keep an off-diagonal light quark mass 
matrix: then, in principle, the light states (in the weak basis) may mix into each 
other under modular transformation, but $V_{\rm CKM}$ and $J_{CP}$ are invariant.

\end{document}